\documentclass[prd,aps,preprint,nofootinbib,amssymb,eqsecnum,showkeys]{revtex4}
\usepackage{verbatim,graphics,graphicx,color,slashed}
\usepackage{ulem} 
\usepackage{hyperref}
\usepackage{changepage}
\usepackage{color}

\def\lsim{\mathrel {\vcenter {\baselineskip 0pt \kern 0pt
    \hbox{$<$} \kern 0pt \hbox{$\sim$} }}}
\def\gsim{\mathrel {\vcenter {\baselineskip 0pt \kern 0pt
    \hbox{$>$} \kern 0pt \hbox{$\sim$} }}}
\def\slashchar#1{\setbox0=\hbox{$#1$}           
 \dimen0=\wd0                                 
  \setbox1=\hbox{/} \dimen1=\wd1               
\ifdim\dimen0>\dimen1                        
  \rlap{\hbox to \dimen0{\hfil/\hfil}}      
  #1                                        
  \else                                        
 \rlap{\hbox to \dimen1{\hfil$#1$\hfil}}   
   /                                         
  \fi}                                         %
\def\cpto{\mathrel {\vcenter {\baselineskip 0pt \kern 0pt
    \hbox{$CP$} \kern 0pt \hbox{$\longrightarrow$} }}}
\def\cptof{\mathrel {\vcenter {\baselineskip 0pt \kern 0pt
    \hbox{$~CP$} \kern 0pt \hbox{$\longleftrightarrow$} }}}

\begin{document}

\baselineskip=15pt

\preprint{}

\title{$SU(3)$ and Isospin Breaking Effects
on $B \to PPP$ Amplitudes }

\author{Xiao-Gang He$^{1,2,3}$\footnote{hexg@phys.ntu.edu.tw}}
\author{Guan-Nan Li${}^{2}$\footnote{lgn198741@126.com}}
\author{Dong Xu$^{1}$\footnote{xudong1104@gmail.com}}

\affiliation{${}^{1}$INPAC, SKLPPC and Department of Physics,
Shanghai Jiao Tong University, Shanghai, China}
\affiliation{${}^{2}$CTS, CASTS and
Department of Physics, National Taiwan University, Taipei, Taiwan}
\affiliation{${}^{3}$Physics Division, National Center for
Theoretical Sciences, Department of Physics, National Tsing Hua
University, Hsinchu, Taiwan}

\date{\today $\vphantom{\bigg|_{\bigg|}^|}$}

\date{\today}

\vskip 1cm
\begin{abstract}
Several modes of $B$ decays into three pseudoscalar octet mesons PPP have been measured. These decays have provided useful information for B decays in the standard model (SM). Some of powerful tools in analyzing B decays are flavor $SU(3)$ and isospin symmetries. Such analyses are usually hampered by $SU(3)$ breaking effects due to a relatively large strange quark mass which breaks SU(3) symmetry down to isospin symmetry. The isospin symmetry also breaks down when up and down quark  mass difference is non-zero. It is therefore interesting to find relations which are not sensitive to $SU(3)$ and isospin breaking effects.  We find that the relations among several fully-symmetric $B \to PPP$ decay amplitudes are not affected by first order $SU(3)$ breaking effects due to a non-zero strange quark mass, and also some of them are not affected by  first isospin breaking effects.   These relations, therefore, hold to good precisions. Measurements for these relations can provide important information about B decays in the SM.
\end{abstract}

\pacs{PACS numbers: }

\maketitle

\section{Introduction}

Several decay modes of $B$ decays into three pseudoscalar octet mesons PPP have been measured\cite{3p-data-babar,3p-data-lhcb}. $B\to PPP$ has been a subject of theoretical studies\cite{he3}. The new data have raised new interests in related theoretical studies\cite{yadong,gronau,xu-li-he1,he-li-xu2,hy-cheng,gronau-full}. With more data from LHCb, one can expect that the study of $B\to PPP$ will provide more important information for B decays in the standard model (SM).

A powerful tool to analyze B decays is flavor $SU(3)$ symmetry\cite{su31}.
Some of the interesting features of using flavor $SU(3)$ are the predictions of relations among different decay modes which can be experimentally tested. The flavor $SU(3)$ symmetry is, however, expected to be only an approximate symmetry because $u$, $d$ and $s$ quarks have different masses.  Since the strange quark has a relative larger mass compared with those of up and down quarks, it is the larger source of symmetry breaking. If up and down quark masses are neglected, a non-zero strange quark mass breaks flavor $SU(3)$ symmetry down to the isospin symmetry. When up and down quark mass difference is kept, isospin symmetry is also broken. The $SU(3)$ breaking effect is at the level of 20 percent for the $\pi$ and $K$ decay constants $f_\pi$ and $f_K$. For 2-body pseudoscalar octet meson $B$ decays, although there are some $SU(3)$ breakings~\cite{su32}, it works reasonably well, such as rate differences between some of the $\Delta S=0$ and $\Delta S=1$ two-body pseudoscalar meson $B$ decays~\cite{he1,he2}. Analysis has also been carried out for $B\to PPP$ decays using flavor $SU(3)$ recently. It has been shown that the decay and CP asymmetry patterns for the charged $B^+$ decays into $K^+K^-K^+$, $K^+K^-\pi^+$, $K^+\pi^-\pi^+$ and $\pi^+\pi^-\pi^+$ do not follow $SU(3)$ predictions. To explain data, large $SU(3)$ breaking effects are needed\cite{xu-li-he1,he-li-xu2}. Usually isospin breaking effects are much smaller because up and down quark masses are much smaller than the strange quark mass and the QCD scale.

Because of possible large flavor $SU(3)$ breaking effects for $B \to PPP$, the predicted relations among different decay modes can only provide limited information. One wonders whether there exist relations which are immuned from $SU(3)$ or even isospin breaking effects due to $u$, $d$ and $s$ quark mass differences. To this end we carried out an analysis for $B \to PPP$ decays using flavor $SU(3)$ symmetry to identify possible relations, and then include $SU(3)$ breaking effects due to a strange quark mass, and also up and down quark masses to see whether some relations still remain to hold. We find that the relations between several fully-symmetric $B\to PPP$ decay amplitudes studied in Ref. \cite{gronau-full} are not affected by the flavor $SU(3)$ breaking effects due a non-zero strange quark mass, and some of them are not even affected by isospin breaking effects. These relations when measured experimentally can provide useful information about $B$ decays in the SM. In the following we provide some details.

\section{$SU(3)$ conserving amplitudes}

We start with the description of $B$ decays into three pseudoscalar octet mesons from flavor $SU(3)$ symmetry.
The leading quark level effective Hamiltonian up to one loop level in
electroweak interaction for hadronic charmless $B$ decays in the SM can be written as
\begin{eqnarray}
 H_{eff}^q = {4 G_{F} \over \sqrt{2}} [V_{ub}V^{*}_{uq} (c_1 O_1 +c_2 O_2)
   - \sum_{i=3}^{12}(V_{ub}V^{*}_{uq}c_{i}^{uc} +V_{tb}V_{tq}^*
   c_i^{tc})O_{i}],
\end{eqnarray}
where $q$ can be $d$ or $s$the coefficients
$c_{1,2}$ and $c_i^{jk}=c_i^j-c_i^k$, with $j$ and $k$ indicate the internal quark,
are the Wilson Coefficients (WC).  The tree WCs are of order one with, $c_1=-0.31$, and $c_2 = 1.15$. The penguin
WCs are much smaller with the largest one $c_6$ to be $-0.05$. These
WC's have been evaluated by several groups~\cite{heff}. $V_{ij}$ are the KM matrix elements.
In the above the factor $V_{cb}V_{cq}^*$ has
been eliminated using the unitarity property of the KM matrix.

The operators $O_i$ are given by
\begin{eqnarray}
\begin{array}{ll}
O_1=(\bar q_i u_j)_{V-A}(\bar u_i b_j)_{V-A}\;, &
O_2=(\bar q u)_{V-A}(\bar u b)_{V-A}\;,\\
O_{3,5}=(\bar q b)_{V-A} \sum _{q'} (\bar q' q')_{V \mp A}\;,&
O_{4,6}=(\bar q_i b_j)_{V-A} \sum _{q'} (\bar q'_j q'_i)_{V \mp A}\;,\\
O_{7,9}={ 3 \over 2} (\bar q b)_{V-A} \sum _{q'} e_{q'} (\bar q' q')_{V \pm A}\;,\hspace{0.3in} &
O_{8,10}={ 3 \over 2} (\bar q_i b_j)_{V-A} \sum _{q'} e_{q'} (\bar q'_j q'_i)_{V \pm A}\;,\\
O_{11}={g_s\over 16\pi^2}\bar q \sigma_{\mu\nu} G^{\mu\nu} (1+\gamma_5)b\;,&
O_{12}={Q_b e\over 16\pi^2}\bar q \sigma_{\mu\nu} F^{\mu\nu} (1+\gamma_5)b.
\end{array}
\end{eqnarray}
where $(\bar a b)_{V-A} = \bar a \gamma_\mu (1-\gamma_5) b$, $G^{\mu\nu}$ and
$F^{\mu\nu}$ are the field strengths of the gluon and photon, respectively.

At the hadron level, the decay amplitude can be generically written as
\begin{eqnarray}
A = \langle final\;state|H_{eff}^q|\bar {B}\rangle = V_{ub}V^*_{uq} T(q) + V_{tb}V^*_{tq}P(q)\;,
\end{eqnarray}
where $T(q)$ contains contributions from the $tree$ as well as $penguin$ due to charm and up
quark loop corrections to the matrix elements,
while $P(q)$ contains contributions purely from
one loop $penguin$ contributions. $B$ indicates one of the $B^+$, ${B}^0$ and ${B}^0_s$. $B_i = (B^+,  B^0,  B^0_s)$ forms an $SU(3)$ triplet.

The flavor $SU(3)$ symmetry transformation properties for operators $O_{1,2}$, $O_{3-6, 11,12}$, and $O_{7-10}$ are: $\bar 3_a + \bar 3_b +6 + \overline {15}$,
$\bar 3$, and $\bar 3_a + \bar 3_b +6 + \overline {15}$, respectively.
We indicate these representations by matrices in $SU(3)$ flavor space by $H(\bar 3)$, $H(6)$ and $H(\overline{15})$.
For $q=d$, the non-zero entries of the matrices $H(i)$ are given by~\cite{he1}
\begin{eqnarray}
H(\bar 3)^2 &=& 1\;,\;\;
H(6)^{12}_1 = H(6)^{23}_3 = 1\;,\;\;H(6)^{21}_1 = H(6)^{32}_3 =
-1\;,\nonumber\\
H(\overline {15} )^{12}_1 &=& H(\overline {15} )^{21}_1 = 3\;,\; H(\overline
{15} )^{22}_2 =
-2\;,\;
H(\overline {15} )^{32}_3 = H(\overline {15} )^{23}_3 = -1\;.
\end{eqnarray}
And for $q = s$, the non-zero entries are
\begin{eqnarray}
H(\bar 3)^3 &=& 1\;,\;\;
H(6)^{13}_1 = H(6)^{32}_2 = 1\;,\;\;H(6)^{31}_1 = H(6)^{23}_2 =
-1\;,\nonumber\\
H(\overline {15} )^{13}_1 &=& H(\overline {15} ) ^{31}_1 = 3\;,\; H(\overline
{15} )^{33}_3 =
-2\;,\;
H(\overline {15} )^{32}_2 = H(\overline {15} )^{23}_2 = -1\;.
\end{eqnarray}

These properties enable one to
write the decay amplitudes for $B \to PPP$ decays in only a few $SU(3)$ invariant amplitudes~\cite{su31}. Here $P$ is one of the mesons in the pseudoscalar octet meson $M=(M_{ij})$ which is given by,
\begin{eqnarray}
M= \left ( \begin{array}{ccc}
{\pi^0\over \sqrt{2}} + {\eta_8\over \sqrt{6}}&\pi^+&K^+\\
\pi^-&-{\pi^0\over \sqrt{2}} + {\eta_8\over \sqrt{6}}& K^0\\
K^-&\bar K^0&-{2\eta_8\over \sqrt{6}}
\end{array} \right ).
\end{eqnarray}

Construction of $B \to PPP$ decay amplitude can be done order by order by using three $M$'s, $B$, and the Hamiltonian $H$, and also derivatives on the mesons to form $SU(3)$. The $SU(3)$ conserving
momentum independent amplitudes can be constructed by the following.

For the $T(q)$ amplitude, we have\cite{xu-li-he1}
\begin{eqnarray}
T(q)&&=a^T(\overline{3})B_{i}H^{i}(\overline{3})M^{j}_{k}M^{k}_{l}M^{l}_{j}+b^T(\overline{3})H^{i}(\overline{3})M^{j}_{i}B_{j}M^{k}_{l}M^{l}_{k} +c^T(\overline{3})H^{i}(\overline{3})M^{l}_{i}M^{j}_{l}M^{k}_{j}B_{k}\nonumber\\
&&+a^T(6)B_{i}H^{ij}_{k}(6)M^{k}_{j}M^{l}_{n}M^{n}_{l}+b^T(6)B_{i}H^{ij}_{k}(6)M^{k}_{l}M^{l}_{n}M^{n}_{j}\nonumber\\
&&+c^T(6)B_{i}H^{jk}_{l}(6)M^{i}_{j}M^{n}_{k}M^{l}_{n}+d^T(6)B_{i}H^{jk}_{l}(6)M^{i}_{n}M^{l}_{j}M^{n}_{k}\nonumber\\
&&+a^T(\overline {15})B_{i}H^{ij}_{k}(\overline{15})M^{k}_{j}M^{l}_{n}M^{n}_{l}+b^T(\overline{15})B_{i}H^{ij}_{k}(\overline{15})M^{k}_{l}M^{l}_{n}M^{n}_{j}\nonumber\\
&&+c^T(\overline{15})B_{i}H^{jk}_{l}(\overline{15})M^{i}_{j}M^{n}_{k}M^{l}_{n}+d^T(\overline{15})B_{i}H^{jk}_{l}(\overline{15})M^{i}_{n}M^{l}_{j}M^{n}_{k}\;.\label{su3s}
\end{eqnarray}
One can write similar amplitude $P(q)$ for the penguin contributions.

The coefficients $a(i)$, $b(i)$, $c(i)$ and $d(i)$ are constants which contain the WCs and information about QCD dynamics. Expanding the above $T(q)$ amplitude, one can extract the decay amplitudes for specific decays in terms of these coefficients.

In the above we have described how to obtain flavor $SU(3)$ amplitudes which are momentum independent.
However, due to the three body decay nature, in general, there are momentum dependence in the decay amplitudes. The momentum dependence can in principle be determined by analysing Dalitz plots for the decays.
The lowest order terms with derivatives lead to two powers of momentum dependence.
One can obtain relevant terms by taking two times of derivatives on each of the terms in Eq.(\ref{su3s}) and  then collecting them together. It has been shown\cite{xu-li-he1} that there are six independent ways of taking derivatives for each of the terms listed in eq. (\ref{su3s}). For example after taking derivatives for $B_i H^i(\overline 3) M^j_k M^k_l M^l_j$, we have the following independent terms
\begin{eqnarray}
&&(\partial_\mu B_i) H^i(\overline 3) (\partial^\mu M^j_k) M^k_l M^l_j,\;\;(\partial_\mu B_i) H^i(\overline 3) M^j_k (\partial^\mu M^k_l) M^l_j,\;\;(\partial_\mu B_i) H^i(\overline 3) M^j_k M^k_l (\partial^\mu M^l_j)\;,\nonumber\\
&&B_i H^i(\overline 3) (\partial_\mu M^j_k) (\partial^\mu M^k_l) M^l_j,\;\;B_i H^i(\overline 3) (\partial_\mu M^j_k ) M^k_l (\partial^\mu M^l_j),\;\;B_i H^i(\overline 3) M^j_k (\partial_\mu M^k_l) (\partial^\mu M^l_j)\;.\nonumber\\
\end{eqnarray}
The full list of the possible terms have been obtained in Ref.\cite{xu-li-he1} in the Appendix B. We will not repeat them here.

Using the above $SU(3)$ decay amplitudes, one can find some interesting relations among different decays\cite{xu-li-he1}.
It has been recently pointed out  that there are additional relations among the fully-symmetric final states B decay amplitudes ${\cal A}_{FS}$\cite{gronau-full}. Study of these relations can provide further information about flavor $SU(3)$ symmetry in $B$ decays.

The fully-symmetric $B \to PPP$ amplitudes ${\cal A}_{FS}$ is related to the usual decay amplitudes $A(P_1(p_1) P_2(p_2) P_3(p_3))$ for the final mesons $P_{1,2,3}$ carrying momenta $p_{1,2,3}$, for all three final mesons are distinctive, by
\begin{eqnarray}
{\cal A}_{FS}(P_1 &P_2&P_3) = \\
&{1\over \sqrt{3}}& \left ( A(P_1(p_1) P_2(p_2) P_3(p_3)) + A(P_1(p_2) P_2(p_3) P_3(p_1))+A(P_1(p_3) P_2(p_1) P_3(p_2)) \right )\;,\nonumber
\end{eqnarray}
For the cases that two of them or all three of them are identical particles, the identical particle factorial factors should be taken cared. In Ref.\cite{gronau-full}, how the fully-symmetric amplitudes can be determined experimentally has been discussed in detail. We will not repeat the discussions here. We concentrate on how these amplitudes are derived in the framework of flavor SU(3) symmetry and how they are affected by SU(3) breaking effects due to finite quark masses for u, d and s quarks.

To understand that why there are new relations between the fully-symmetric amplitudes for different decay modes, let us consider $B^+\rightarrow K^0\pi^+\pi^0$ and $B^0_d\rightarrow K^+\pi^0\pi^-$ decays as examples.

Expanding eq. (\ref{su3s}),  one obtains
\begin{eqnarray}
T(B^{+} \to K^{0}\pi^{+}\pi^{0})&&={\sqrt{2}}\left ( c(6)+d(6)+{2}c(\overline{15})+{2}d(\overline{15})\right )\;,
\end{eqnarray}
and
\begin{eqnarray}
T(B^{0} \to K^{+}\pi^{-}\pi^{0}) = T(B^{+} \to K^{0}\pi^{+}\pi^{0})\;.
\end{eqnarray}
From which we get $T_{FS}(B^{+} \to K^{0}\pi^{+}\pi^{0})=T_{FS}(B^{0} \to K^{+}\pi^{-}\pi^{0})$.

As the decay amplitudes may have momentum dependence, we should also check if the equality of the above two amplitudes are equal when taking into account of momentum dependence in the amplitudes. Expanding terms in Appendix B of Ref.\cite{xu-li-he1}, we find
\begin{eqnarray}
&&T^p(B^{+} \to K^{0}\pi^{+}\pi^{0})=\nonumber\\
&&\qquad\qquad\alpha_1p_{B}\cdot p_{1}+\alpha_2p_{B}\cdot p_{2}+\alpha_3p_{B}\cdot p_{3}+\alpha_4p_{1}\cdot p_{2}+\alpha_5p_{1}\cdot p_{3}+\alpha_6p_{2}\cdot p_{3}\;,\nonumber\\
&&T^p(B^{0}_d \to K^{+}\pi^{0}\pi^{-})=\nonumber\\
&&\qquad\qquad\beta_1p_{B}\cdot p_{1}+\beta_2p_{B}\cdot p_{2}+\beta_3p_{B}\cdot p_{3}+\beta_4p_{1}\cdot p_{2}+\beta_5p_{1}\cdot p_{3}+\beta_6p_{2}\cdot p_{3}\;.
\end{eqnarray}

The coefficients $\alpha_i$ and $\beta_i$ are given by,
\begin{eqnarray}
\alpha_1&=&\sqrt{2}\left ( c^{\prime}(6)_{2}+2c^{\prime}(\overline{15})_{2}+d^{\prime}(6)_{3}+2d^{\prime}(\overline{15})_{3}\right )\;,\nonumber\\
\alpha_2&=&\frac{1}{\sqrt{2}}\left (-b^{\prime}(6)_{1}+b^{\prime}(6)_{2}-3 b^{\prime}(\overline{15})_{1}+3b^{\prime}(\overline{15})_{2}+c^{\prime}(\overline{3})_{2}-c^{\prime}(\overline{3})_{3} +c^{\prime}(6)_{1}\right .\nonumber\\&+&\left . c^{\prime}(6)_{3}+c^{\prime}(\overline{15})_{1}+3c^{\prime}(\overline{15})_{3}+d^{\prime}(6)_{1}+d^{\prime}(6)_{3} +5d^{\prime}(\overline{15})_{1}-d^{\prime}(\overline{15})_{3} \right )\;,\nonumber\\
\alpha_3&=&\frac{1}{\sqrt{2}}\left (b^{\prime}(6)_{1}-b^{\prime}(6)_{2}+3b^{\prime}(\overline{15})_{1}-3b^{\prime}(\overline{15})_{2}-c^{\prime}(\overline{3})_{2}+c^{\prime}(\overline{3})_{3}+c^{\prime}(6)_{1}+c^{\prime}(6)_{3} \right .\\&+&\left . 3c^{\prime}(\overline{15})_{1}+c^{\prime}(\overline{15})_{3}+d^{\prime}(6)_{1}+2d^{\prime}(6)_{2}-d^{\prime}(6)_{3}-d^{\prime}(\overline{15})_{1}+4 d^{\prime}(\overline{15})_{2}+d^{\prime}(\overline{15})_{3}\right )\;,\nonumber\\
\alpha_4&=&\frac{1}{\sqrt{2}}\left (- b^{\prime\prime}(6)_{2}+b^{\prime\prime}(6)_{3}-3b^{\prime\prime}(\overline{15})_{2}+3b^{\prime\prime}(\overline{15})_{3}+c^{\prime\prime}(\overline{3})_{1}-c^{\prime\prime}(\overline{3})_{2}+c^{\prime\prime}(6)_{1}+c^{\prime\prime}(6)_{3}\right .\nonumber\\
&+&\left . c^{\prime\prime}(\overline{15})_{1}+3c^{\prime\prime}(\overline{15})_{2}-d^{\prime\prime}(6)_{1}+2d^{\prime\prime}(6)_{2}+d^{\prime\prime}(6)_{3}+d^{\prime\prime}(\overline{15})_{1}-d^{\prime\prime}(\overline{15})_{2}+4d^{\prime\prime}(\overline{15})_{3}\right )\;,\nonumber\\
\alpha_5&=&\frac{1}{\sqrt{2}}\left (b^{\prime\prime}(6)_{2}-b^{\prime\prime}(6)_{3}+3b^{\prime\prime}(\overline{15})_{2}-3b^{\prime\prime}(\overline{15})_{3}-c^{\prime\prime}(\overline{3})_{1}+c^{\prime\prime}(\overline{3})_{2}+c^{\prime\prime}(6)_{1}\right .\nonumber\\
&+&\left . c^{\prime\prime}(6)_{3}+3c^{\prime\prime}(\overline{15})_{1}+c^{\prime\prime}(\overline{15})_{2}+d^{\prime\prime}(6)_{1}+d^{\prime\prime}(6)_{3}-d^{\prime\prime}(\overline{15})_{1}+d^{\prime\prime}(\overline{15})_{2}\right )\;,\nonumber\\
\alpha_6&=&\sqrt{2}\left ( c^{\prime\prime}(6)_{2}+2 c^{\prime\prime}(\overline{15})_{3}+d^{\prime\prime}(6)_{1}+2d^{\prime\prime}(\overline{15})_{1}\right )\;. \nonumber
\end{eqnarray}
and
 \begin{eqnarray}
\beta_1&=&\sqrt{2}\left (c^{\prime}(6)_{2}+2c^{\prime}(\overline{15})_{2}+d^{\prime}(6)_{3}+2d^{\prime}(\overline{15})_{3}\right )\;,\nonumber\\
\beta_2&=&\frac{1}{\sqrt{2}}\left (b^{\prime}(6)_{1}-b^{\prime}(6)_{2}+b^{\prime}(\overline{15})_{1}-b^{\prime}(\overline{15})_{2}+c^{\prime}(\overline{3})_{2}-c^{\prime}(\overline{3})_{3}+c^{\prime}(6)_{1}+c^{\prime}(6)_{3}\right .\nonumber\\
&+&\left .c^{\prime}(\overline{15})_{1}+ 3c^{\prime}(\overline{15})_{3}+d^{\prime}(6)_{1}+2d^{\prime}(6)_{2}-d^{\prime}(6)_{3}-3d^{\prime}(\overline{15})_{1}+4d^{\prime}(\overline{15})_{2}+3d^{\prime}(\overline{15})_{3}\right )\;,\nonumber\\
\beta_3&=&\frac{1}{\sqrt{2}}\left ( -b^{\prime}(6)_{1}+b^{\prime}(6)_{2}-b^{\prime}(\overline{15})_{1}+b^{\prime}(\overline{15})_{2}-c^{\prime}(\overline{3})_{2}+c^{\prime}(\overline{3})_{3}+c^{\prime}(6)_{1}\right .\\
&+&\left . c^{\prime}(6)_{3}+3c^{\prime}(\overline{15})_{1}+c^{\prime}(\overline{15})_{3}+d^{\prime}(6)_{1}+d^{\prime}(6)_{3}+7 d^{\prime}(\overline{15})_{1}-3d^{\prime}(\overline{15})_{3}\right )\;,\nonumber\\
\beta_4&=&\frac{1}{\sqrt{2}}\left (b^{\prime\prime}(6)_{2}-b^{\prime\prime}(6)_{3}+b^{\prime\prime}(\overline{15})_{2}-b^{\prime\prime}(\overline{15})_{3}+c^{\prime\prime}(\overline{3})_{1}-c^{\prime\prime}(\overline{3})_{2}+c^{\prime\prime}(6)_{1}\right .\nonumber\\
&+& \left . c^{\prime\prime}(6)_{3}+c^{\prime\prime}(\overline{15})_{1}+3c^{\prime\prime}(\overline{15})_{2}+d^{\prime\prime}(6)_{1}+d^{\prime\prime}(6)_{3}-3d^{\prime\prime}(\overline{15})_{1}+7d^{\prime\prime}(\overline{15})_{2}\right )\;,\nonumber\\
\beta_5&=&\frac{1}{\sqrt{2}}\left ( - b^{\prime\prime}(6)_{2}+b^{\prime\prime}(6)_{3}-b^{\prime\prime}(\overline{15})_{2}+b^{\prime\prime}(\overline{15})_{3}-c^{\prime\prime}(\overline{3})_{1}+c^{\prime\prime}(\overline{3})_{2}+c^{\prime\prime}(6)_{1}+ c^{\prime\prime}(6)_{3}\right .\nonumber\\
&+&\left . 3c^{\prime\prime}(\overline{15})_{1}+c^{\prime\prime}(\overline{15})_{2}-d^{\prime\prime}(6)_{1}+2 d^{\prime\prime}(6)_{2}+d^{\prime\prime}(6)_{3}+3d^{\prime\prime}(\overline{15})_{1}-3d^{\prime\prime}(\overline{15})_{2}+4d^{\prime\prime}(\overline{15})_{3}\right )\;,\nonumber\\
\beta_6&=&\sqrt{2}\left (c^{\prime\prime}(6)_{2}+2c^{\prime\prime}(\overline{15})_{3}+d^{\prime\prime}(6)_{1}+2d^{\prime\prime}(\overline{15})_{1}\right )\;.\nonumber
\end{eqnarray}

One can see from the above that $T^p(B^{+} \to K^{0}\pi^{+}\pi^{0})$ is no longer equal to $T^p(B^{0}_d \to K^{+}\pi^{0}\pi^{-})$. However, one can readily see from the above equations, that
\begin{eqnarray}
\alpha_1+\alpha_2+\alpha_3=\beta_1+\beta_2+\beta_3,~~\alpha_4+\alpha_5+\alpha_6=\beta_4+\beta_5+\beta_6\;.
\end{eqnarray}
This fact makes the fully-symmetric amplitudes to satisfy
\begin{eqnarray}
T^p(B^{+} \to K^{0}\pi^{+}\pi^{0})_{FS} = T^p(B^{0}_d \to K^{+}\pi^{0}\pi^{-})_{FS}\;.
\end{eqnarray}
Similarly, the penguin amplitudes $P$ and $P^p$ have the same properties discussed above for the tree amplitudes, $T$ and $T^p$.

The total fully-symmetric amplitudes ${\cal A}_{FS} = V_{ub} V^*_{uq} (T_{FS} + T^p_{FS}) + V_{tb}V_{tq}^* (P_{FS} + P^p_{FS})$ then have the relation
\begin{eqnarray}
{\cal A}(B^{+} \to K^{0}\pi^{+}\pi^{0})_{FS} = {\cal A}(B^{0}_d \to K^{+}\pi^{0}\pi^{-})_{FS}\;.
\end{eqnarray}
Enlarging the amplitudes to fully-symmetric ones, indeed produce more relations.

Expanding eq. (\ref{su3s}) and equations in Appendix B of Ref. \cite{xu-li-he1}, we obtain the following relations confirming those obtained in Ref.\cite{gronau-full}.
For $\bar b \to \bar s$ induced $B\to PPP$ amplitudes, we have
\\

$ 1.~B\to K\pi\pi$
\\

$S1.1 = {\cal A}(B^+\to K^0\pi^+\pi^0)_{\rm FS} - {\cal A}(B^0 \to K^+\pi^0\pi^-)_{\rm FS} =0$,

$S1.2 = \sqrt{2}{\cal A}(B^+\to K^0\pi^+\pi^0)_{\rm FS} - {\cal A}(B^0 \to K^0\pi^+\pi^-)_{\rm FS} +  2{\cal A}(B^0\to K^0\pi^0\pi^0)_{\rm FS} =0$,

$S1.3 = \sqrt{2}{\cal A}(B^0\to K^+\pi^0\pi^-)_{\rm FS} + {\cal A}(B^+\to K^+\pi^+\pi^-)_{\rm FS} -  2{\cal A}(B^+\to K^+\pi^0\pi^0)_{\rm FS} =0$.
\\

$2.~B\to KK{\bar K}$
\\

$S2.1 = - {\cal A}(B^+\to K^+K^+K^-)_{\rm FS} + {\cal A}(B^+\to K^+K^0{\bar K}^0)_{\rm FS} $

$\;\;\;\;\;\;\;\;+  {\cal A}(B^0 \to K^0K^+K^-)_{\rm FS} - {\cal A}(B^0 \to K^0 K^0{\bar K}^0)_{\rm FS} =0$.
\\

$3.~B^0_{s}\to \pi K{\bar K}$
\\

$S3.1 = \sqrt{2}{\cal A}(B^0_s\to \pi^0 K^+K^-)_{\rm FS} - \sqrt{2}{\cal A}(B^0_s\to \pi^0  K^0{\bar K}^0)_{\rm FS} $

$\;\;\;\;\;\;\;\;-~ {\cal A}(B^0_s\to \pi^- K^+ {\bar K}^0)_{\rm FS} - {\cal A}(B^0_s\to \pi^+ K^-K^0)_{\rm FS} =0 $.
\\

$4.~B^0_{s}\to \pi\pi\pi$
\\

$S4.1 =  2{\cal A}(B^0_s\to \pi^0\pi^0\pi^0)_{\rm FS} - {\cal A}(B^0_s \to \pi^0\pi^+\pi^-)_{\rm FS}=0 $.
\\

For $\bar b \to \bar d$ induced  $B\to PPP$ amplitudes, we have
\\

$1.~B\to \pi K{\bar K}$
\\

$D1.1 = -\sqrt{2} {\cal A}(B^0 \to \pi^0 K^+K^-)_{\rm FS}
+ {\cal A}(B^0\to \pi^+ K^0K^-)_{\rm FS}
- {\cal A}(B^+ \to \pi^+ K^+K^-)_{\rm FS}$

$\;\;\;\;\;\;\;\;+~\sqrt{2} {\cal A}(B^0 \to \pi^0 K^0{\bar K}^0)_{\rm FS}
+ {\cal A}(B^0 \to \pi^- K^+{\bar K}^0)_{\rm FS} $

$\;\;\;\;\;\;\;\; +~{\cal A}(B^+ \to \pi^+ K^0{\bar K}^0)_{\rm FS}
- \sqrt{2} {\cal A}(B^+ \to \pi^0 K^+{\bar K}^0)_{\rm FS} = 0 $.
\\

$2.~B \to \pi\pi\pi$
\\

$D2.1 =  2 {\cal A}(B^0\to \pi^0\pi^0\pi^0)_{\rm FS} -  {\cal A}(B^0\to \pi^+\pi^0\pi^-)_{\rm FS}=0 $,

$D2.2 = 2 {\cal A}(B^+\to \pi^+\pi^0\pi^0)_{\rm FS} - {\cal A}(B^+\to \pi^-\pi^+\pi^+)_{\rm FS}=0 $.
\\

$3.~B^0_{s}\to K\pi\pi$
\\

$D3.1 = -2{\cal A}(B^0_s\to {\bar K}^0\pi^0\pi^0)_{\rm FS} + {\cal A}(B^0_s\to{\bar K}^0\pi^+\pi^-)_{\rm FS} - \sqrt{2}{\cal A}(B^0_s\to K^-\pi^+\pi^0)_{\rm FS} = 0 $.
\\

In the above, we have considered some relations among decay processes with the same $\Delta S$. There are also some other relations among tree and penguin amplitudes but with different $\Delta S$. Some of them will be discussed later in the conclusions.

Note that we have different normalizations than those used in Ref.\cite{gronau-full} for some of the final meson states and also identical particle combinatorial factors. One can easily obtain relations in the form in Ref. \cite{gronau-full} by
multiplying a ``-1'' to the amplitudes when $\pi^{-},K^{-},\pi^{0}$ appear each time as one of the final states, and a factor $1/\sqrt{2}$ and  $1/\sqrt{6}$ in our formulation for the corresponding amplitudes, respectively, when the decays involve two and three identical particles.

\section{$SU(3)$ and Isospin breaking due to quark mass differences}

The main source for flavor $SU(3)$ symmetry breaking effects comes from difference in
masses of $u$, $d$ and $s$ quarks. Under $SU(3)$, the mass matrix
can be viewed as combinations of representations from $3\times \bar
3$, to matching the ($u$, $d$, $s$) transformation property as a
fundamental representation, which contains an 1 and an 8 irreducible
representations. The diagonalized  mass matrix can be expressed as a
linear combination of the identity matrix $I$, and the Gell-Mann
matrices $\lambda_3$ and $\lambda_8$. We have
\begin{eqnarray}
\left ( \begin{array}{ccc}
 m_u & 0 & 0\\
 0 & m_d& 0 \\
0 &  0 & m_s
\end{array}
\right ) = {1\over 3}(m_u+m_d+m_s) I + {1\over 2}(m_u - m_d) X + {1\over 6}(m_u+m_d - 2 m_s) W\;,
\end{eqnarray}
with $X$ and $W$ given by
\begin{eqnarray}
X=\left ( \begin{array}{ccc}
 1 & 0 & 0\\
 0 & -1& 0 \\
0 &  0 & 0
\end{array}
\right )\;,\;\;\;\;W=\left ( \begin{array}{ccc}
 1 & 0 & 0\\
 0 & 1& 0 \\
0 &  0 & -2
\end{array}
\right )\;.
\end{eqnarray}

Compared with $s$-quark mass
$m_s$, the $u$ and $d$ quark masses $m_{u,d}$  are much smaller, $SU(3)$ breaking effects due to a non-zero $m_s$
dominates the $SU(3)$ breaking effects. When up and down quark mass difference is neglected, the residual symmetry of $SU(3)$ becomes the isospin symmetry. In that case when studying $SU(3)$ breaking effects,
the term proportional to
$X$ can be dropped. The identity $I$ part contributes to the $B$ decay amplitudes in a similar way as that given in eq. (\ref{su3s}) which can be absorbed into the coefficients $a(i)$ to $d(i)$. Only $W$ piece will contribute to the $SU(3)$ breaking effects. We will first discuss this case to  first order in $W$, and then also study the isospin breaking effects by including the first order term proportional to $X$.

\subsection{SU(3) breaking due a non-zero $m_s$}

To construct relevant decay amplitudes for $B\to PPP$ decays, one first breaks the contraction of indices at any joint in eq. (\ref{su3s}), and inserts a W in between,  and then contracts all indices appropriately. For example corresponding to the first term in eq. (\ref{su3s}), there are two ways to insert $W$,
\begin{eqnarray}
B_{i}H^{a}(\overline{3})W^{i}_{a}M^{j}_{k}M^{k}_{l}M^{l}_{j}\;,\;\;\;\;B_{i}H^{i}(\overline{3})M^{j}_{k}M^{k}_{l}M^{a}_{j}W^{l}_{a}\;.
\end{eqnarray}
The full list of possible independent terms are given in Appendix A of Ref.\cite{xu-li-he1}.

Extracting the $SU(3)$ breaking terms for $B^{+} \to K^{0}\pi^{+}\pi^{0}$ and $B^{0} \to K^{+}\pi^{0}\pi^{-}$ decays, we have the corrections for the decay amplitudes, $\Delta T$, as
\begin{eqnarray}
\triangle T(B^{+} &\to& K^{0}\pi^{+}\pi^{0}
) =\sqrt{2}\left ( c^T_{1}(6)+\sqrt{2}c^T_{2}(6)-2c^T_{3}(6)+\sqrt{2}c^T_{4}(6)+c^T_{5}(6)+c^T_{1}(\overline{15})\right .\nonumber\\&+&\left . c^T_{2}(\overline{15})-2c^T_{3}(\overline{15})+ c^T_{4}(\overline{15})+c^T_{5}(\overline{15})+d^T_{1}(6)+d^T_{2}(6)-2d^T_{3}(6)+d^T_{4}(6)\right .\nonumber\\
&+&\left . d^T_{5}(6)+d^T_{1}(\overline{15})+d^T_{2}(\overline{15})-2d^T_{3}(\overline{15})+d^T_{4}(\overline{15})+d^T_{5}(\overline{15})\right )\;,
\end{eqnarray}
and
\begin{eqnarray}
\triangle T(B^{0} \to K^{+}\pi^{0}\pi^{-})=\triangle T(B^{+} \to K^{0}\pi^{+}\pi^{0})\;,
\end{eqnarray}
which leads to the equality of the fully-symmetric amplitudes for these two decays.
Therefore,
\begin{eqnarray}
S1.1 = {\cal A}(B^{0} \to K^{+}\pi^{0}\pi^{-})_{FS}-{\cal A}(B^{+} \to K^{0}\pi^{+}\pi^{0})_{FS}=0\;,
\end{eqnarray}
still holds.

Note that even $SU(3)$ breaking effects affect each of the decay amplitudes, the relation of the fully-symmetric amplitudes of these two decays are not affected {by the first order $SU(3)$ breaking effects.}
Expanding terms in Appendix A of Ref.\cite{xu-li-he1}, one can study relations discussed above. We find that all the relations among the fully-symmetric amplitudes still hold, that is
\begin{eqnarray}
&&S1.1 = 0\;,\;\;S2.1 = 0\;,\;\;S1.3=0\;\,\;\;S2.1=0\;,\;\;S3.1=0\;,\;\;S4.1=0\;,\nonumber\\
&&D1.1=0\;,\;\;D2.1=0\;,\;\;D2.2=0\;,\;\;D3.1=0\;,\label{resu3}
\end{eqnarray}
are still true even if one include $SU(3)$ breaking effects due a non-zero strange quark mass.
This actually is not a surprise because the relations discussed can be obtained by isospin symmetry considerations.

Experimental verification of these relations may provide important tests for the validity of flavor $SU(3)$ for $B$ decays.

\subsection{Isospin breaking due to up and down quark mass difference}

It would be interesting to investigate what happens when mass difference between up and down quark, which breaks isospin symmetry, is also included. We now discuss these isospin breaking effects for the relations discussed before.

One can obtain the corrections by replacing $W$ by $X$ in Appendix A of Ref.\cite{xu-li-he1}. We indicate the coefficients in a similar way as that $SU(3)$ breaking effects due to a non-zero $m_s$, but with a superscript $I$ to indicate the effects of isospin breaking, for example for tree operator corrections by $a^{T^I}_i$, $b^{T^I}_i$, $c^{T^I}_i$, and $d^{T^I}_i$. The correction to the decay amplitude will also be indicated by a superscript $I$, $\Delta T^I$.

Expanding all terms, we obtain the corrections due to isospin breaking effects for all the decay amplitudes discussed previously. We find that except that the relation $S4.1$ still holds, all other relations for the $B\to PPP$ decay amplitudes induced by $\bar b \to \bar s$ and $\bar b\to \bar d$ interactions discussed earlier are broken.

In fact each of the decay modes relevant in S4.1 is affected by isospin breaking effects,
\begin{eqnarray}
&&\Delta T^{I}(B^0_{s}\rightarrow \pi^0\pi^0\pi^0)=\nonumber\\
&&\qquad\quad\frac{\sqrt{2}}{2}(a^{T^{I}}_{2}(\overline{3})+2a^{T^{I}}_{2}(\overline{15})+2a^{T^{I}}_{3}(\overline{15})+b^{T^{I}}_{2}(\overline{15})+b^{T^{I}}_{3}(\overline{15})+b^{T^{I}}_{4}(\overline{15})+b^{T^{I}}_{5}(\overline{15}))\;,\nonumber\\
&&\Delta T^{I}(B^0_{s}\rightarrow \pi^+\pi^-\pi^0)=\\
&&\qquad\quad\sqrt{2}(a^{T^{I}}_{2}(\overline{3})+2a^{T^{I}}_{2}(\overline{15})+2a^{T^{I}}_{3}(\overline{15})+b^{T^{I}}_{2}(\overline{15})+b^{T^{I}}_{3}(\overline{15})+b^{T^{I}}_{4}(\overline{15})+b^{T^{I}}_{5}(\overline{15}))\;,\nonumber
\end{eqnarray}
but they are affected in such a way that the equality of the amplitudes is not affected. That is, we still have:
\begin{eqnarray}
S4.1=2 {\cal A}(B^0_s\to \pi^0\pi^0\pi^0)_{\rm FS} - {\cal A}(B^0_s \to \pi^0\pi^+\pi^-)_{\rm FS}=0.\label{reis1}
\end{eqnarray}

 This makes this relation special because that this relation is
not affected by first order $SU(3)$ breaking effects due to a non-zero strange quark and isospin breaking due to up and down quark mass difference. It should hold to a high precision. Experimental test of this relation can provide important information about $B \to PPP$.

We also found some other interesting relations even isospin violating effects are included, namely the corrections for some of the relations discussed above are related to others.
For $b\to s$ interaction induced decay modes,
we have an additional relation which relate $S1.2$ and $S1.3$ because the isospin breaking effects satisfy
\begin{eqnarray}
&&[\sqrt{2}\Delta T^{I}(B^+\rightarrow K^0\pi^+\pi^0)-\Delta T^{I}(B^0\rightarrow K^0\pi^+\pi^-)+2\Delta T^{I}(B^0\rightarrow K^0\pi^0\pi^0)]\nonumber\\
&&=-[\sqrt{2}\Delta T^{I}(B^0\rightarrow K^+\pi^-\pi^0)+\Delta T^{I}(B^+\rightarrow K^+\pi^+\pi^-)-2\Delta T^{I}(B^+\rightarrow K^+\pi^0\pi^0)]\;.\nonumber\\
\end{eqnarray}

Although the right hand sides of $S1.2$ and $S1.3$ are not zero anymore, the above relation leads to,
\begin{eqnarray}
S1.2 =-S1.3\neq 0\;.\label{reis2}
\end{eqnarray}

For $b\to d$ interactions induced decay modes, we have
\begin{eqnarray}
&&[2\Delta T^{I}(B^+\rightarrow \pi^0\pi^0\pi^+)-\Delta T^{I}(B^+\rightarrow \pi^-\pi^+\pi^+)]\nonumber\\
&&=-[-2\Delta T^{I}(B^0_{s}\rightarrow \bar{K^0}\pi^0\pi^0)+\Delta T^{I}(B^0_{s}\rightarrow \bar{K^0}\pi^+\pi^-)-\sqrt{2}\Delta T^{I}(B^0_{s}\rightarrow K^-\pi^+\pi^0)]\;,\nonumber\\
&&\sqrt{2}[2\Delta T^{I}(B^0\rightarrow \pi^0\pi^0\pi^0)-\Delta T^{I}(B^0\rightarrow \pi^+\pi^-\pi^0)]\\
&&=-[2\Delta T^{I}(B^+\rightarrow \pi^0\pi^0\pi^+)-\Delta T^{I}(B^+\rightarrow \pi^-\pi^+\pi^+)]\;.\nonumber
\end{eqnarray}

Due to isospin breaking effects, the right hand sides of $D2.1$, $D2.2$ and $D3.1$ are non-zero. However, the above relations imply
\begin{eqnarray}
\sqrt{2} D2.1 = -D2.2\neq 0\;,\;\;\;\;D2.2= -D3.1 \neq 0\;.\label{reis3}
\end{eqnarray}

We would like to emphasize that since the above relations hold even when first order isospin effects have been taken into account, they can provide useful information about $B$ decays in the SM in a way independent of flavor $SU(3)$ and isospin breaking effects.

\subsection{Momentum dependent $SU(3)$ and isospin breaking amplitudes}

 There are also momentum dependent amplitudes at the same order to the $SU(3)$ and isospin breaking effects discussed in the previous subsections. We find that all the relations eq. (\ref{resu3}),  and eqs. (\ref{reis1}), (\ref{reis2}) and (\ref{reis3}) still hold when $SU(3)$ breaking effects due to a non-zero $m_s$, and isospin breaking effects due to $m_u$ and $m_d$ mass difference discussed earlier, respectively.
The analysis is similar to the case for $SU(3)$ conserving momentum amplitudes. We will not give details here, but just outline how the analysis can be carried out. The leading ones are constructed by taking two powers of derivatives on each term of the $SU(3)$ breaking amplitudes which have been shown in the Appendix A of Ref.\cite{xu-li-he1}. For example for the term, $a^T_{1}(\overline{3}) B_{i}H^{a}(\overline{3})W^{i}_{a}M^{j}_{k}M^{k}_{l}M^{l}_{j}$, we obtain the following six independent terms with two derivatives:
\begin{eqnarray}
&&a^{\prime}_{1}(\overline{3})_{1}(\partial_\mu B_{i}) H^{a}(\overline{3})W^{i}_{a} (\partial_\mu M^{j}_{k}) M^{k}_{l}M^{l}_{j},\;\;
a^{\prime}_{1}(\overline{3})_{2}(\partial_\mu B_{i}) H^{a}(\overline{3})W^{i}_{a}M^{j}_{k} (\partial_\mu M^{k}_{l}) M^{l}_{j},\;\nonumber\\
&&a^{\prime}_{1}(\overline{3})_{3}(\partial_\mu B_{i} ) H^{a}(\overline{3})W^{i}_{a}M^{j}_{k}M^{k}_{l}  (\partial_\mu M^{l}_{j}), \;\;
a^{\prime\prime}_{1}(\overline{3})_{1} B_{i}H^{a}(\overline{3}) W^{i}_{a}(\partial_\mu M^{j}_{k} ) (\partial_\mu M^{k}_{l}) M^{l}_{j},\;\nonumber\\
&& a^{\prime\prime}_{1}(\overline{3})_{2}B_{i}H^{a}(\overline{3}) W^{i}_{a}(\partial_\mu M^{j}_{k} )  M^{k}_{l} (\partial_\mu  M^{l}_{j}),\;\;
 a^{\prime\prime}_{1}(\overline{3})_{3} B_{i}H^{a}(\overline{3}) W^{i}_{a} M^{j}_{k}  (\partial_\mu M^{k}_{l})(\partial_\mu M^{l}_{j}).
\end{eqnarray}
Here, $a^{\prime}_{1}(\overline{3})_{1}, a^{\prime}_{1}(\overline{3})_{2}, a^{\prime}_{1}(\overline{3})_{3}, a^{\prime\prime}_{1}(\overline{3})_{1}, a^{\prime\prime}_{1}(\overline{3})_{2}, a^{\prime\prime}_{1}(\overline{3})_{3}$ are constants. We then extend similar definition of constants for other $SU(3)$ breaking terms in the Appendix A and B of Ref.\cite{xu-li-he1}. There are both tree and penguin amplitudes which can be further labeled by superscripts $T$ and $P$. We will omit writing them out with the understanding that what described below work for both tree and penguin amplitudes.
One obtains the relevant terms for isospin breaking effects by replacing $W$ by $X$.

Expanding all possible terms, one obtains the amplitudes. Taking the amplitudes in $S1.1$ for illustration, one finds that the momentum dependent amplitudes can be written as
\begin{eqnarray}
&&\Delta T^p(B^{+} \to K^{0}\pi^{+}\pi^{0})=\nonumber\\
&&\qquad\qquad \Delta\alpha_1p_{B}\cdot p_{1}+\Delta\alpha_2p_{B}\cdot p_{2}+\Delta\alpha_3p_{B}\cdot p_{3}+\Delta\alpha_4p_{1}\cdot p_{2}+\Delta\alpha_5p_{1}\cdot p_{3}+\Delta\alpha_6p_{2}\cdot p_{3}\;,\nonumber\\
&& \Delta T^p(B^{0}_d \to K^{+}\pi^{0}\pi^{-})=\\
&&\qquad\qquad\Delta\beta_1p_{B}\cdot p_{1}+\Delta\beta_2p_{B}\cdot p_{2}+\Delta\beta_3p_{B}\cdot p_{3}+\Delta\beta_4p_{1}\cdot p_{2}+\Delta\beta_5p_{1}\cdot p_{3}+\Delta\beta_6p_{2}\cdot p_{3}\;.\nonumber
\end{eqnarray}

The coefficients $\Delta\alpha_{i},\Delta\beta_{i}$ are collections of coefficients from all possible terms.
Our detailed calculations show that
\begin{eqnarray}
&&\Delta\alpha_1+\Delta\alpha_2+\Delta\alpha_3=\Delta\beta_1+\Delta\beta_2+\Delta\beta_3\nonumber\\
&&=\sqrt{2}[2c^{\prime}_{1}(\overline {15})_{1}+2c^{\prime}_{1}(\overline {15})_{2}+2c^{\prime}_{1}(\overline {15})_{3}+
2c^{\prime}_{2}(\overline {15})_{1}+2c^{\prime}_{2}(\overline {15})_{2}+2c^{\prime}_{2}(\overline {15})_{3}-
4c^{\prime}_{3}(\overline {15})_{1}\nonumber\\&&-4c^{\prime}_{3}(\overline {15})_{2}
-4c^{\prime}_{3}(\overline {15})_{3}+2c^{\prime}_{4}(\overline {15})_{1}+2c^{\prime}_{4}(\overline {15})_{2}+2c^{\prime}_{4}(\overline {15})_{3}+2c^{\prime}_{5}(\overline {15})_{1}+2c^{\prime}_{5}(\overline {15})_{2}+2c^{\prime}_{5}(\overline {15})_{3}\nonumber\\&&+c^{\prime}_{1}(6)_{1}+c^{\prime}_{1}(6)_{2}
+c^{\prime}_{1}(6)_{3}+c^{\prime}_{2}(6)_{1}+c^{\prime}_{2}(6)_{2}+c^{\prime}_{2}(6)_{3}-2c^{\prime}_{3}(6)_{1}-2c^{\prime}_{3}(6)_{2}
-2c^{\prime}_{3}(6)_{3}+c^{\prime}_{4}(6)_{1}\nonumber\\&&+c^{\prime}_{4}(6)_{2}+c^{\prime}_{4}(6)_{3}
+c^{\prime}_{5}(6)_{1}+c^{\prime}_{5}(6)_{2}+c^{\prime}_{5}(6)_{3}+2d^{\prime}_{1}(\overline {15})_{1}+2d^{\prime}_{1}(\overline {15})_{2}+2d^{\prime}_{1}(\overline {15})_{3}+2d^{\prime}_{2}(\overline {15})_{1}\nonumber\\&&+2d^{\prime}_{2}(\overline {15})_{2}+
2d^{\prime}_{2}(\overline {15})_{3}
-4d^{\prime}_{3}(\overline {15})_{1}-4d^{\prime}_{3}(\overline {15})_{2}-
4d^{\prime}_{3}(\overline {15})_{3}+2d^{\prime}_{4}(\overline {15})_{1}+2d^{\prime}_{4}(\overline {15})_{2}+
2d^{\prime}_{4}(\overline {15})_{3}\nonumber\\&&+2d^{\prime}_{5}(\overline {15})_{1}+2d^{\prime}_{5}(\overline {15})_{2}
+2d^{\prime}_{5}(\overline {15})_{3}
+d^{\prime}_{1}(6)_{1}+d^{\prime}_{1}(6)_{2}+d^{\prime}_{1}(6)_{3}+
+d^{\prime}_{2}(6)_{1}+d^{\prime}_{2}(6)_{2}+d^{\prime}_{2}(6)_{3}\nonumber\\&&-2d^{\prime}_{3}(6)_{1}-2d^{\prime}_{3}(6)_{2}-2d^{\prime}_{3}(6)_{3}
+d^{\prime}_{4}(6)_{1}+d^{\prime}_{4}(6)_{2}+d^{\prime}_{4}(6)_{3}+d^{\prime}_{5}(6)_{1}+d^{\prime}_{5}(6)_{2}+d^{\prime}_{5}(6)_{3}]\nonumber\\
&&\Delta\alpha_4+\Delta\alpha_5+\Delta\alpha_6=\Delta\beta_4+\Delta\beta_5+\Delta\beta_6\nonumber\\
&&=\sqrt{2}[2c^{\prime\prime}_{1}(\overline {15})_{1}+2c^{\prime\prime}_{1}(\overline {15})_{2}+2c^{\prime\prime}_{1}(\overline {15})_{3}+
2c^{\prime\prime}_{2}(\overline {15})_{1}+2c^{\prime\prime}_{2}(\overline {15})_{2}+2c^{\prime\prime}_{2}(\overline {15})_{3}-
4c^{\prime\prime}_{3}(\overline {15})_{1}\nonumber\\&&-4c^{\prime\prime}_{3}(\overline {15})_{2}
-4c^{\prime\prime}_{3}(\overline {15})_{3}+2c^{\prime\prime}_{4}(\overline {15})_{1}+2c^{\prime\prime}_{4}(\overline {15})_{2}+2c^{\prime\prime}_{4}(\overline {15})_{3}+2c^{\prime\prime}_{5}(\overline {15})_{1}+2c^{\prime\prime}_{5}(\overline {15})_{2}+2c^{\prime\prime}_{5}(\overline {15})_{3}\nonumber\\&&+c^{\prime\prime}_{1}(6)_{1}+c^{\prime\prime}_{1}(6)_{2}
+c^{\prime\prime}_{1}(6)_{3}+c^{\prime\prime}_{2}(6)_{1}+c^{\prime\prime}_{2}(6)_{2}+c^{\prime\prime}_{2}(6)_{3}-2c^{\prime\prime}_{3}(6)_{1}-2c^{\prime\prime}_{3}(6)_{2}
-2c^{\prime\prime}_{3}(6)_{3}+c^{\prime\prime}_{4}(6)_{1}\nonumber\\&&+c^{\prime\prime}_{4}(6)_{2}+c^{\prime\prime}_{4}(6)_{3}
+c^{\prime\prime}_{5}(6)_{1}+c^{\prime\prime}_{5}(6)_{2}+c^{\prime\prime}_{5}(6)_{3}+2d^{\prime\prime}_{1}(\overline {15})_{1}+2d^{\prime\prime}_{1}(\overline {15})_{2}+2d^{\prime\prime}_{1}(\overline {15})_{3}+2d^{\prime\prime}_{2}(\overline {15})_{1}\nonumber\\&&+2d^{\prime\prime}_{2}(\overline {15})_{2}+
2d^{\prime\prime}_{2}(\overline {15})_{3}
-4d^{\prime\prime}_{3}(\overline {15})_{1}-4d^{\prime\prime}_{3}(\overline {15})_{2}-
4d^{\prime\prime}_{3}(\overline {15})_{3}+2d^{\prime\prime}_{4}(\overline {15})_{1}+2d^{\prime\prime}_{4}(\overline {15})_{2}+
2d^{\prime\prime}_{4}(\overline {15})_{3}\nonumber\\&&+2d^{\prime\prime}_{5}(\overline {15})_{1}+2d^{\prime\prime}_{5}(\overline {15})_{2}
+2d^{\prime\prime}_{5}(\overline {15})_{3}
+d^{\prime\prime}_{1}(6)_{1}+d^{\prime\prime}_{1}(6)_{2}+d^{\prime\prime}_{1}(6)_{3}+
+d^{\prime\prime}_{2}(6)_{1}+d^{\prime\prime}_{2}(6)_{2}+d^{\prime\prime}_{2}(6)_{3}\nonumber\\&&-2d^{\prime\prime}_{3}(6)_{1}-2d^{\prime\prime}_{3}(6)_{2}-2d^{\prime\prime}_{3}(6)_{3}
+d^{\prime\prime}_{4}(6)_{1}+d^{\prime\prime}_{4}(6)_{2}+d^{\prime\prime}_{4}(6)_{3}+d^{\prime\prime}_{5}(6)_{1}+d^{\prime\prime}_{5}(6)_{2}+d^{\prime\prime}_{5}(6)_{3}]\nonumber
\end{eqnarray}
With these facts, after symmetrizing the amplitude to the fully-symmetric one, we find $S1.1 =0$ still holds.
We find that the other relations of eq.(\ref{resu3}) also hold.

In a very similar way one can obtain the momentum dependent corrections to the isospin breaking effects by replacing $W$ by $X$ as what have been done for $SU(3)$ case. We find that all the relations of eq.(\ref{reis1}), eq.(\ref{reis2}) and eq.(\ref{reis3}) still hold.

 Before close this section, we would like to make a comment about finite mass effects of $m^2_\pi$ and $m^2_K$. In practical extraction of the amplitudes, one should also consider $SU(3)$ corrections in phase space due to final state meson mass differences which come in order $m^2_{\pi, K}/m^2_{B, B_s}$ since $m^2_\pi \sim m_u, m_d$ and $m^2_K \sim m_s$ which are the same order of $SU(3)$ and isospin breaking effects considered earlier. This can be done systematically when extracting the fully-symmetric amplitudes by Dalitz plot analysis. In the momentum dependent amplitudes discussed in section II, when express the amplitudes, for example those in Eq.(2.12), in terms of the $s$, $t$ and $u$ variables, terms proportional to $m^2_\pi$ and $m_K^2$ will be generated. However, these will not generate new terms compared with those already included in the $SU(3)$ and isospin breaking effects considered earlier in this section. The conclusions drawn above will not be changed.

\section{Conclusions and Discussions}

Flavor $SU(3)$ and isospin symmetries have been considered to be powerful tools in analyzing B decays. Such analyses are usually hampered by a relatively large strange quark mass which breaks $SU(3)$ symmetry down to isospin symmetry. The isospin symmetry also breaks down when up and down quark  mass difference is kept. It is therefore interesting to find relations which are not sensitive to $SU(3)$ and isospin breaking effects.  We have carried out detailed analyses including $SU(3)$ and isospin breaking effects due to u, d and s quark mass differences for $B \to PPP$ decays. We find that a class of relations in fully-symmetric amplitudes are not broken by $SU(3)$ breaking effects due to a non-zero strange quark mass, and
the relations
\begin{eqnarray}
S4.1=0\;,\;\;S1.2+S1.3 = 0\;,\;\;\sqrt{2}D2.1+D2.2=0\;,\;\;D2.2+D3.1=0\;,
\end{eqnarray}
hold even isospin breaking effects due to up and down quark mass difference is included.
Measurements for these relations will provide important information about B decays in the SM.

We would like to end the paper by commenting $SU(3)$ breaking effects on the $U$-spin symmetry relations in the following
\begin{eqnarray}
&&T_{\triangle s=-1}(B^+\rightarrow K^+K^+K^-)=T_{\triangle s=0}(B^+\rightarrow \pi^+\pi^+\pi^-)\;,\nonumber\\
&&T_{\triangle s=-1}(B^+\rightarrow K^+\pi^+\pi^-)=T_{\triangle s=0}(B^+\rightarrow \pi^+K^+K^-)\;.
\label{uspin}
\end{eqnarray}
The momentum dependent terms also respect the above relations. The above equalities also hold for the fully-symmetric amplitudes for corresponding pairs of decay modes in the $SU(3)$ limit.
These relations imply in the SM that the CP violating rate asymmetries defined by
$A_{asy} = \Gamma(B\to PPP) - \Gamma(\bar B \to \bar P\bar P\bar P)$ are equal but opposite in sign for each pair of decay modes above.

For the fully-symmetric amplitudes of these decays modes, we also have
\begin{eqnarray}
&&T_{\triangle s=-1}(B^+\rightarrow K^+K^+K^-)_{FS}=T_{\triangle s=-1}(B^+\rightarrow K^+\pi^+\pi^-)_{FS}\;,\nonumber\\
&&T_{\triangle s=0}(B^+\rightarrow \pi^+\pi^+\pi^-)_{FS}=T_{\triangle s=0}(B^+\rightarrow \pi^+K^+K^-)_{FS}\;.
\label{uspin-fs}
\end{eqnarray}

Unlike the other fully-symmetric amplitudes studied in previous sections, the relations in eq.(\ref{uspin}) and eq.(\ref{uspin-fs}) are broken when $SU(3)$ breaking effects due to a non-zero strange quark mass  is included. Therefore there may be sizeable  deviation for these relations.
Relations in eq.(\ref{uspin}) have been discussed recently. It was found that indeed there are large $SU(3)$ breaking effects\cite{xu-li-he1, he-li-xu2,gronau}. The relations in eq. (\ref{uspin}) and eq.(\ref{uspin-fs}) will not provide as much insight as those from the fully-symmetric amplitudes which still hold when isospin breaking effects are included discussed earlier.

However, we find that the $SU(3)$ breaking effects due to a non-zero strange quark mass and the isospin breaking effects due to the difference of up and down quark masses are equal for some of the above relations with
\begin{eqnarray}
&&\Delta T(B^{+}\rightarrow K^+ \pi^+ \pi^-)-\Delta T(B^{+}\rightarrow K^+ K^+ K^- )\nonumber\\
&& =- [\Delta T(B^{+}\rightarrow \pi^+ K^+ K^-)-\Delta T(B^{+}\rightarrow \pi^+ \pi^+ \pi^- )]\\
&&=3[a^T_{4}(6)+3a^T_{4}(\overline{15})+b^T_{3}(\overline{3})+b^T_{4}(6)+3b^T_{4}(\overline{15})+c^T_{2}(\overline{3})-c^T_{2}(6)+c^T_{3}(6)\nonumber\\
&&\quad-c^T_{5}(6)-c^T_{2}(\overline{15})-c^T_{3}(\overline{15})-2c^T_{4}(\overline{15})+3c^T_{5}(\overline{15})-d^T_{5}(6)+3d^T_{5}(\overline{15})]\;,\nonumber
\end{eqnarray}
and the isospin breaking effects satisfy
\begin{eqnarray}
&&\Delta T^{I}(B^{+}\rightarrow K^+ \pi^+ \pi^-)-\Delta T^{I}(B^{+}\rightarrow K^+ K^+ K^- )\nonumber\\
&&=- [\Delta T^{I}(B^{+}\rightarrow \pi^+ K^+ K^-)-\Delta T^{I}(B^{+}\rightarrow \pi^+ \pi^+ \pi^- )]\\
&&=-[a^{T^{I}}_{4}(6)+3a^{T^{I}}_{4}(\overline{15})+b^{T^{I}}_{3}(\overline{3})+b^{T^{I}}_{4}(6)+3b^{T^{I}}_{4}(\overline{15})+c^{T^{I}}_{2}(\overline{3})-c^{T^{I}}_{2}(6)+c^{T^{I}}_{3}(6)\nonumber\\
&&\quad-c^{T^{I}}_{5}(6)-c^{T^{I}}_{2}(\overline{15})-c^{T^{I}}_{3}(\overline{15})-2c^{T^{I}}_{4}(\overline{15})+3c^{T^{I}}_{5}(\overline{15})-d^{T^{I}}_{5}(6)+3d^{T^{I}}_{5}(\overline{15})].\nonumber
\end{eqnarray}

 With the  momentum dependent corrections to the $SU(3)$ and isospin breaking effects, detailed analyses similar to those carried out in Sectin III, we find that the following relation is still true to the order we can considering
\begin{eqnarray}
&&\Delta T^{p}(B^{+}\rightarrow K^+ \pi^+ \pi^-)-\Delta T^{p}(B^{+}\rightarrow K^+ K^+ K^- )\nonumber\\
&& =- [\Delta T^{p}(B^{+}\rightarrow \pi^+ K^+ K^-)-\Delta T^{p}(B^{+}\rightarrow \pi^+ \pi^+ \pi^- )],
\end{eqnarray}
Here $\Delta T^{p}$ including both the momentum dependent corrections to the $SU(3)$ and isospin breaking effects.

The above leads to the following relation which is not affected by first order $SU(3)$ breaking effects due to strange, up and down quark mass differences,
\begin{eqnarray}
&&T (B^{+}\rightarrow K^+ \pi^+ \pi^- )_{FS}-T(B^{+}\rightarrow K^+ K^+ K^- )_{FS} \nonumber\\
&&=T(B^{+}\rightarrow \pi^+ \pi^+ \pi^-)_{FS} - T(B^{+}\rightarrow \pi^+ K^+ K^- )_{FS}\neq 0\;,
\end{eqnarray}
and similarly for penguin amplitudes $P_{FS}$.

When the relevant decay amplitudes are measured precisely, one can also obtain useful information for $B$ decays in the SM.

\begin{acknowledgments}

The work was supported in part by MOE Academic Excellent Program (Grant No: 102R891505) and MOST of ROC, and in part by NNSF(Grant No:11175115) and Shanghai Science and Technology Commission (Grant No: 11DZ2260700) of PRC.

\end{acknowledgments}

\end{document}